\documentclass[prl
,twocolumn%
,floatfix%
,superscriptaddress%
,showpacs]{revtex4}
\usepackage{bm}
\usepackage[T1]{fontenc}
\usepackage[latin1]{inputenc}

\usepackage{graphicx}%
\usepackage{dcolumn}
\usepackage{amsmath}
\usepackage{pstricks}

\makeatletter
\def\btt#1{\texttt{\@backslashchar#1}}%
\DeclareRobustCommand\bblash{\btt{\@backslashchar}}%
\makeatother

\newcommand{\doot}{\mathaccent"17 }
\newlength{\Mi}
\newlength{\Mii}
\newcommand{\Vol}[1]{{\bf #1}}
\newcommand{\DL}{{\ensuremath{\delta}-layer}}
\newcommand{\CoCu}{{\ensuremath{Co_9/Cu_7}}}
\newcommand{\FeCr}{{\ensuremath{Fe_9/Cr_7}}}
\newcommand{\FeAu}{{\ensuremath{Fe_9/Au_7}}}
\newcommand{\EV}[1]{(#1)}
\renewcommand{\EV}[1]{}
\newcommand{\Eq}[1]{\mbox{Eq.~(\ref{#1})}}
\newcommand{\Fig}[1]{\mbox{Fig.~\ref{#1}}}
\newcommand{\FIG}[1]{\mbox{Fig.~{#1}}}
\newcommand{\Mcite}[1]{\mbox{Ref.~\cite{#1}}}

\newcommand{\bM}[1]{\begin{minipage}[t]{#1}}
\newcommand{\eM}{\end{minipage}}
\begin{document}
\bibliographystyle{apsrev}
\title{
Impurity scattering and quantum confinement in giant magnetoresistance
systems
}
\author{Peter Zahn}%
\email{Zahn@physik.uni-halle.de}
\affiliation{%
Fachbereich Physik, Martin-Luther-Universit{\"a}t Halle-Wittenberg,
D-06099 Halle, Germany
}%
%
\author{J\"org Binder}%
\affiliation{%
Institut f\"ur Theoretische Physik, Technische Universit\"at Dresden,
D-01062 Dresden, Germany
}%
\altaffiliation{present address:
Heyde AG, Auguste-Viktoria-Str. 2, D-61231 Bad Nauheim, Germany}
\author{Ingrid Mertig}%
\affiliation{%
Fachbereich Physik, Martin-Luther-Universit{\"a}t Halle-Wittenberg,
D-06099 Halle, Germany
}%
%
\date{\today}
\begin{abstract}
Ab initio calculations for the giant magnetoresistance 
(GMR) in Co/Cu, Fe/Cr, and Fe/Au multilayers are presented. The 
electronic structure of the multilayers 
and the scattering potentials of point defects therein are 
calculated self-consistently. Residual resistivities are 
obtained by solving the quasi-classical Boltzmann equation 
including the electronic structure of the layered system,
the anisotropic scattering cross sections derived by a Green's function
method and the 
vertex corrections. 
Furthermore, the influence of scattering centers at the interfaces and
within the metallic layers is 
incorporated by averaging the scattering cross sections of 
different impurities at various sites. 
An excellent agreement of 
experimental and theoretical results concerning the general 
trend of GMR in Co/Cu systems depending on the type and the 
position of impurities is obtained. 
Due to the quantum confinement in magnetic 
multilayers GMR can be tailored as a function of the impurity
position. In Co/Cu and Fe/Au systems impurities in the magnetic layer lead to high GMR
values, whereas in Fe/Cr systems defects at the interfaces are most
efficient to increase GMR.
\end{abstract}
%
\maketitle
%
%
\section{introduction}
A large number of experimental and
theoretical investigations was initiated by 
the discovery of giant magnetoresistance (GMR) in magnetic multilayers
\cite{baibich88,binasch89} to elucidate the microscopic origin of the
phenomenon. 
Several authors have shown \cite{tsymbal01a,mertig02} that
GMR in magnetic multilayers is 
strongly influenced by changes in the electronic structure%
, especially the Fermi velocities, 
of the system in dependence on the relative
magnetization alignment in adjacent layers.
In realistic samples, however, spin-dependent scattering is considered
to cause GMR. The effort to tailor GMR systems with high rates was
  accompanied by a variety of experiments \cite{parkin93b,marrows01} and 
calculations \cite{levy94,zahn98,blaas99a} which investigated the
influence of dusting and doping by impurities. Agreement was reached
concerning the dominant role of interface scattering
\cite{parkin93b}. The results of Marrows et al. \cite{marrows01}, however,
demonstrated the strong dependence of GMR on the position of 
the impurities with respect to the interfaces and on the valence
difference between impurity and host.
The aim of this paper is to present ab initio calculations for the
scattering cross sections and resulting GMR ratios in dependence on
defect material and position for the standard systems of
magnetoelectronics Co/Cu and Fe/Cr. 
In addition, the system Fe/Au was
investigated because the system shows a high interface quality
\cite{unguris97} and the
structural properties should be closest in experiment and theory.  
The influence of defects on interlayer exchange coupling in Fe/Au was
investigated earlier \cite{opitz01}. 
\section{Method}
All calculations are performed within the framework of density
functional theory 
in local spin density approximation applying
the Screened KKR (Korringa-Kohn-Rostoker) Green's function
method \cite{zeller95}. 
We have chosen a multilayer geometry in the
so-called first antiferromagnetic (AF) maximum of interlayer exchange coupling for
the Co/Cu system 
consisting of 9 monolayers (ML) Co separated by 7 ML Cu, denoted as
Co$_9$Cu$_7$. 
The structure of the superlattice was assumed to be an fcc lattice with
a lattice constant of 6.76 a.u. grown in (001) direction.
Each atomic plane is represented by one atom in the prolonged unit
cell with 32 atoms.
A similar configuration was chosen for Fe/Cr and a perfect bcc
stacking in (001) direction with a lattice constant of $5.50 a.u.$ was
assumed. 
This is samewhat larger than the lattice constant of Fe and Cr, but provides the
correct magnetic order within the LDA and the atomic sphere approximation for
the potentials \cite{moruzzi92}.
The Fe/Cr system consists of 9 ML Fe and 9 ML Cr which possesses an AF ground
state confirmed by calculation and experiment \cite{unguris97}.
Structural relaxations at the interfaces were neglected.
Despite of the AF order the Cr layer will be refered to as the
nonmagnetic layer in the system, to use the same notation as for the other
system under consideration.
We consider Fe/Au multilayers with the same structural data reported by the
experiments \cite{unguris97} that is a structural bcc-fcc transition. 
The lattice constant for Fe is $a_{bcc}=5.4163 a.u.$, and for Au
$a_{fcc}=\sqrt{2} a_{bcc}$. The thickness of 9 ML of the Au layer was chosen
in accordance with the experimentally obtained interlayer exchange coupling
strength which favors an antiparallel coupling for this Au thickness.

The self-consistent electronic structure of the ideal host, without any
impurities, is described by the one-particle Green's-function, 
whose structural part $\doot{G}^{nn'}_{LL'}(E)$ is 
expanded into a site and
angular-momentum basis 
\cite{zeller79}.

The new aspect for superlattices
is that we consider a lattice with a basis. The index $n$ is now 
a shorthand notation for lattice vector ${\bf R}_N$ and basis vector
${\bf r}_i$ of the atoms in the unit cell.
To simulate a substitutional point defect one atom in the lattice is
replaced by another. The site index of the impurity position is
denoted by $\mu$.

The impurity Green's function ${G}^{nn'} _{LL'}(E)$
of the multilayer including an impurity
atom at a defined position is obtained by the solution of an algebraic
Dyson equation \cite{zeller79}
\begin{equation}\label{eq_dyson}
G^{nn'} _{LL'}(E)
=\doot{G}^{nn'}_{LL'}(E)  +\sum_{n'' L''}
\doot{G}^{nn''} _{LL''}(E) \: \Delta t^{n''}_{L''}(E) \:
{G}^{n''n'}_{L''L'}(E). 
\end{equation}
%
The $n''$-summation involves all sites in the vicinity of the site
$\mu$ for which the differences of the single site t-matrices
$\Delta t^{n''}_{L''}=t^{n''}_{L''}-\doot{t}^{n''}_{L''}$
of the multilayer with and without defect are significant.
The single site t-matrices are derived from the angular-momentum
dependent scattering phase shifts of the potentials in atomic sphere
approximation (ASA). The differences 
$\Delta t_L^n $ characterize the
potential perturbation caused by the defect.
In the calculations we take into account angular momenta $l_{max}\leq 3$.
We allow for potential perturbations up to the second atomic shell
around the impurity atom.
Charge multipoles up to $l_{max}=6$ are taken into account.
Since the systems under consideration are magnetic
all properties mentioned above depend also on spin quantum numbers
$\sigma=\uparrow,\downarrow$
for majority and minority electrons, respectively.

Using the impurity Green function,
the self-consistently calculated potential perturbation 
and the wave function coefficients of the superlattice Bloch states 
we derive the microscopic
spin-conserving transition probability $P^{\sigma}_{kk'}$ for
elastic, that is on-shell-scattering of a Bloch wave $k$ into a
perturbed Bloch wave $k'$. 
$k$ is now a shorthand
notation for the wave vector ${\bf k}$ and band index $\nu$, 
$\sigma$ denotes the spin quantum number
\cite{oppeneer87ab,mertig99}.
The transition probability 
is given by Fermi's golden rule 
\begin{equation}
\label{eq_pkk}
P^{\sigma\sigma}_{kk'} = 2 \pi c N | T^\sigma_{kk'} | ^2 
\delta ( E^\sigma _k - E^\sigma _{k'} ).
\end{equation}
The formalism is restricted to dilute alloys since we assume a linear
dependence 
with the number of defects
$cN$. 
Furthermore, spin-orbit coupling and
the resulting spin-flip
processes are neglected by the non-relativistic scheme.

The transition matrix elements $T^\sigma_{kk'}$ for the
scattering of Bloch electrons by an impurity cluster embedded in an otherwise
ideal translational invariant multilayer are given by
\cite{mertig87,mertig99}
\begin{equation}
\label{eq_tkk}
T^\sigma_{kk'} =  \frac{1}{V}  \sum _{LL'nn'} C^{n}_L (k,\sigma)
T^{nn'} _{LL'} C^{n'} _{L'} (k',\sigma) .
\end{equation}
$C^{n} _L(k,\sigma)$ are the expansion coefficients for the superlattice wave
function in an angular momentum basis. $V$ denotes the total crystal volume.
Using spherical potentials (ASA) the
matrix elements $T^{nn'} _{LL'}$
are derived from the structural Green's function matrix elements 
$G^{nn'}_{LL'}$ of the perturbed system and the 
potential perturbation $\Delta t^n_l$
\cite{oppeneer87a,oppeneer87b,mertig87}.
\EV{
\begin{equation}
\label{eq_tnn}
T^{nn'} _{LL'} = e^{-2 i \doot{\eta}^{\mu}_l} \Delta t^n _L
(1+ G^{nn'} _{L L'} \Delta t^{n'} _{L'})
\end{equation}
Here $ \doot{\eta}^{\mu}_l$ are the scattering phase shifts of the unperturbed
potentials.
}
\par
Summation over all final states
leads to the spin and state dependent relaxation time 
\begin{eqnarray}\label{tau}
\frac{1}{\tau^\sigma_k(\mu)}=\sum_{k'} P^\sigma_{kk'}(\mu)\;
\end{eqnarray}
which depends on spin $\sigma$, Bloch state $k$ and impurity position 
${\bf r}_\mu$ in the superlattice. 
The ${\bf k}$ and spin-dependence of the scattering rates is treated
fully quantum-mechanically without adjustable parameter.
The dependence on the effect position is involved by the implicit dependence
of the impurity Green's function $G^{nn'} _{L L'}$ on the ${\bf r}_\mu$.
Up to this point we consider a dilute alloy of impurity atoms all of them
occupying a chosen site $\mu$ in the unit cell. That is, the alloying
is restricted to certain atomic planes in the multilayer. These
planes correspond to the impurity $\delta$-layers experimentally investigated
by Marrows and Hickey \cite{marrows01}.

The conductivity is calculated by solving the quasi-classical
Boltzmann equation \cite{mertig99}.
Thus, the vector mean free paths are obtained by 
\begin{eqnarray}
{\bf \Lambda}_k^{\sigma}(\mu)=\tau_k^{\sigma}(\mu)\left[
{\bf v}_k^{\sigma} +
\sum\limits_{k'}P_{kk'}^\sigma (\mu) {\bm \Lambda}_{k'}^{\sigma}(\mu)
\right]\;.
\label{eq_boltz}
\end{eqnarray}
This includes besides to the anisotropic relaxation times as the second term on the r.h.s the computational demanding
scattering-in term (vertex corrections) which completes the
description of impurity scattering \cite{swihart86,mertig99}.
The band structure is included via Fermi velocities
${\bf v}_k^{\sigma}$ and the $k'$ summation over
all states on the anisotropic Fermi surface. 
The impurity scattering enters via the
relaxation times $\tau^\sigma_k(\mu)$ and the microscopic
transition probabilities  $P^{\sigma}_{kk'}(\mu)$.
\Eq{eq_boltz} is solved iteratively to determine the vector
mean free path 
${\bm \Lambda}_k^{\sigma}(\mu)$ of an electron with spin $\sigma$ in a
state $k$. 
To our knowledge, up to now the semi-classical calculations of the GMR
have been mostly performed within relaxation time approximation only
\cite{camley89,hood92,zahn95,butler96,zahn98,blaas99a}
, which neglects the scattering-in term.
Zhang and Butler proposed a simplified method to include the vertex
corrections by renormalization of the electron life times using an adjustable
parameter \cite{zhang00}.
The deviation of the relaxation time approximation in comparison with the
solution including the vertex corrections are discussed in \Fig{fig_vertex}.
\par
\begin{figure*}[th]
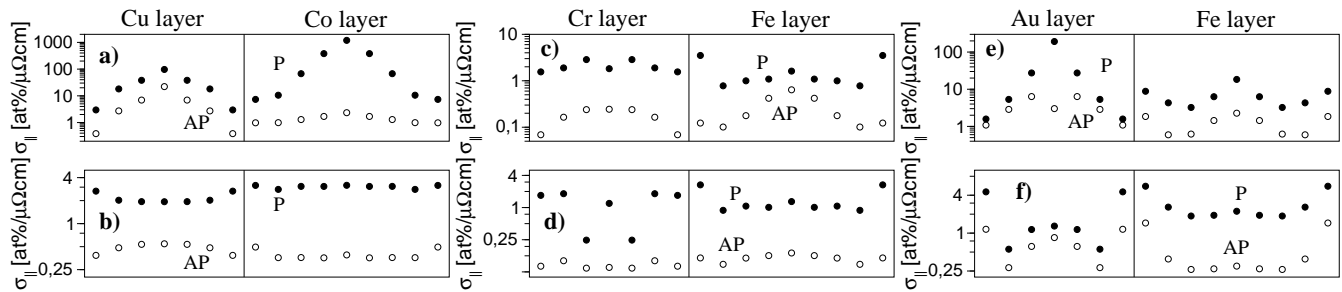

\Mii=.33\linewidth
\Mi=.99\Mii
\bM{\Mii}
\includegraphics[width=\Mi,clip=true]{fig1a.eps}\\
\includegraphics[width=\Mi,clip=true]{fig1b.eps}
\eM
\bM{\Mii}
\includegraphics[width=\Mi,clip=true]{fig1c.eps}\\
\includegraphics[width=\Mi,clip=true]{fig1d.eps}
\eM
\bM{\Mii}
\includegraphics[width=\Mi,clip=true]{fig1e.eps}\\
\includegraphics[width=\Mi,clip=true]{fig1f.eps}
\eM
\caption{
  Conductivity of \CoCu, \FeCr, and {\FeAu} for P and AP alignment in
  dependence on the position of 'self' impurities, respectively;
  a,c,e) Assuming scattering at the inserted $\delta$-layer only,
  b,d,f) assuming $\delta$-layer scattering ($50\%$) and interface
  scattering ($50\%$) 
}
\label{fig_sig_all}
\end{figure*}
Based on the solution of \Eq{eq_boltz} the spin-dependent
conductivity tensor $\sigma^\sigma(\mu)$ 
is given by a Fermi surface integral \cite{mertig99}
\begin{eqnarray}
\label{eq_spincond}
\sigma^\sigma(\mu)=    \frac{e^2}{V}
\sum\limits_k \delta
(E_k^{\sigma}-E_F){\bf \Lambda}_k^{\sigma}(\mu) \circ{\bf v}_k^{\sigma}\; .
\end{eqnarray}
%
With Mott's two current model \cite{mott64} the total conductivity 
$\sigma(\mu)=\sigma^\uparrow(\mu) +\sigma^\downarrow(\mu)$
is obtained by spin summation.
The CIP resistivity is obtained by the inverse of the CIP conductivity due to
the diagonal structure of the conductivity tensor for the tetragonal systems
under consideration
\begin{eqnarray}
\rho(\mu)=    \frac{1}{\sigma^\uparrow(\mu) +
\sigma^\downarrow(\mu)}
\label{eq_resist}
\; .
\end{eqnarray}
To describe the existence of an overall distribution of impurities in
the multilayer the transition probabilities of the different
$\delta$-layers have to be superimposed. Following this idea,
layer-dependent relaxation times $\tau^\sigma_k(\mu)$ are added
%
\begin{eqnarray}
\label{eq_tau_mix}
\frac{1}{\tau^\sigma_k}=\sum_{\mu} \frac{x(\mu)}{\tau^\sigma_k(\mu)}
\;,
\label{eq:tau-mix}
\end{eqnarray}
including weighting factors $x(\mu)$ that account for 
the relative concentration of
defects at the corresponding positions $\mu$ in the unit cell.

The most driving aspect of magnetic multilayers is the drastic change
of the
conductivity $\sigma$ as a function of the relative orientation of the
magnetic layer 
moments, parallel (P) or anti-parallel (AP).
The relative change defines the GMR ratio
\begin{eqnarray}\label{GMR}
GMR(\mu)=\frac{\sigma^{P}(\mu)}{\sigma^{AP}(\mu)}-1
\quad\mbox{.}
\end{eqnarray}
\section{Conductivity and Quantum well states}
We consider the multilayers described above and investigate
the scattering properties of impurities of the nonmagnetic component in the
magnetic layers and vice versa.

%
The analysis of the transport coefficients is focused on the
current-in-plane (CIP) geometry.
The total CIP-conductivities normalized to the defect concentration $c$
caused by impurities of the magnetic material at different positions in the
nonmagnetic layer and
nonmagnetic defects in the magnetic layer are shown for all considered systems
in \Fig{fig_sig_all}a,c,e), respectively, 
for both configurations of the magnetization directions (P, AP).
The conductivity differs by orders of magnitude as a function of
impurity position (keep in mind the logarithmic scale).
The largest values occur for impurity positions
where the quantum confinement produces many
Bloch states with low probability amplitude.
The eigenstates show strong quantum confinement
due to the superlattice potential.
That is, the
probability amplitude is modulated by the layered structure and can even tend
to zero at particular sites of the supercell \cite{butler96,zahn98}.
The consideration of additional interface scattering in \Fig{fig_sig_all}
b,d,f) will be discussed below.
\par
\begin{figure}
  \Mi=.95\linewidth
  \includegraphics[width=\Mi,clip=true]{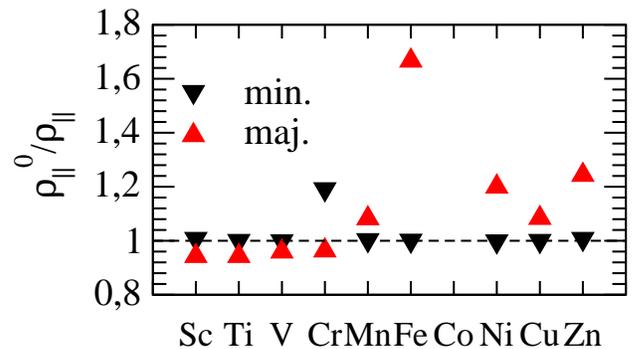}
  \caption{
  Influence of vertex corrections: Dependence of the spin dependent
  resistivity for $Co_9/Cu_7$ multilayers on
  3d impurities at the interface position of the magnetic layer
  $\rho_\parallel$ calculated including vertex corrections and 
  $\doot\rho_\parallel$ without
   }
  \label{fig_vertex}
\end{figure}
The influence of vertex corrections is quantified in \Fig{fig_vertex}. For
different defects at the Co interface position in Co/Cu the deviation of the
resistivity $\rho_\parallel$ calculated including vertex corrections in
\Eq{eq_boltz} and without ($\doot\rho_\parallel$) are shown for both spin
bands. In 
the minority channel (upwards triangles) the deviations are of the order of
few percent (except for Cr). In the majority channel the vertex
corrections are large for defects where sp-scattering dominates. This is the
case for impurities with a similar electronic structure like the host, at least
for this spin direction. For Sc, Ti, V, and Cr with an opposite magnetic moment
with respect to Co and a dominating sd scattering the vertex corrections are small as
for the minority spin direction. Summarizing, one can state, that the variations of
the resistivity by incorporation of the vertex corrections in the solution of
the Boltzmann equation are less than 20\% and do not change the qualitative
behaviour of GMR. The neglect of vertex corrections may change the results
quantitatively, but the general trend is conserved.

\begin{figure}
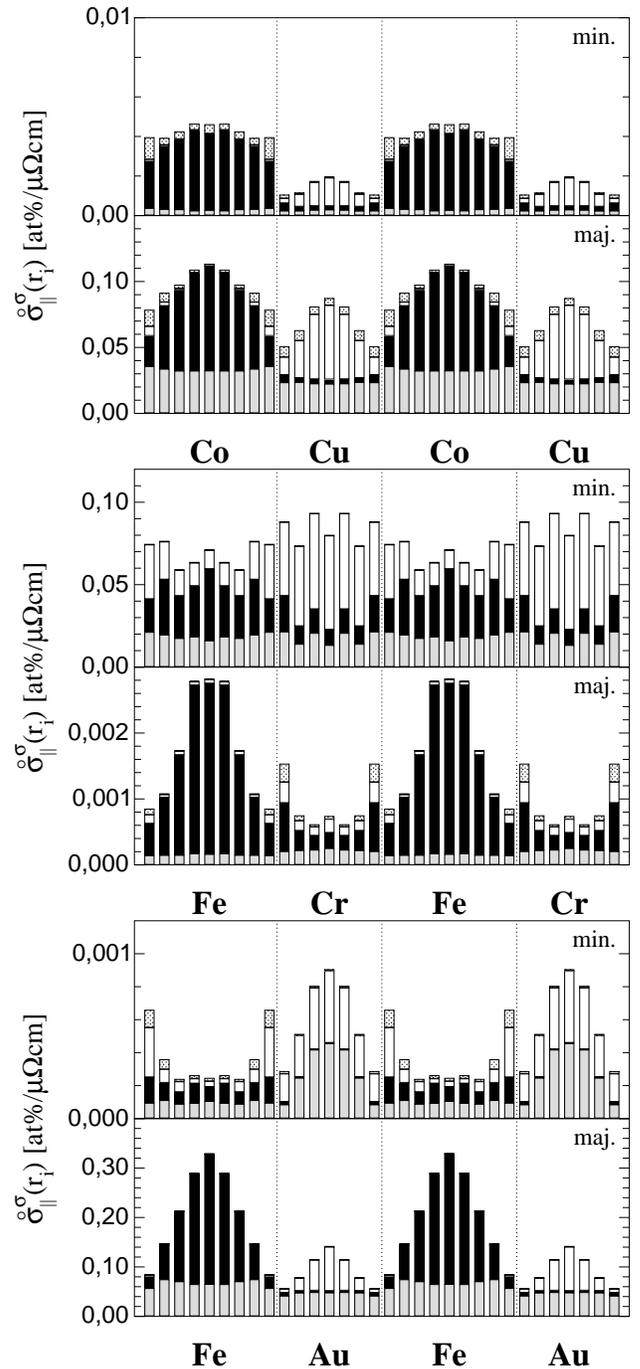

\Mi=.95\linewidth
\includegraphics[width=\Mi,clip=true]{fig5a.eps}\\
\includegraphics[width=\Mi,clip=true]{fig5b.eps}
\includegraphics[width=\Mi,clip=true]{fig5c.eps}
\caption{
  Spin-dependent, layer-projected conductivity of $Co_9Cu_7$,
  $Fe_9Cr_7$, and $Fe_9Au_7$ for P alignment with interface defects,
  typical classes are marked by 
  grey - extended, black - magnetic QWS (Co, Fe), white - nonmagnetic
  QWS (Cu, Cr, Au), and  light grey interface states.
}
\label{fig_sig_local}
\end{figure}

To analyze the influence of the quantum confinement on the character of the
eigenstates a classification of the eigenstates according to \Mcite{zahn98} was
performed. The distribution of the probability amplitude in the different
regions of the multilayer was
analyzed and the states are labeld by the region with the highest averaged
probability amplitude. The eigenstates are classified in 4 types and are
labeled by quantum well states (QWS) in the magnetic layer $C_M$ (Co, Fe), QWS
in the nonmagnetic layer $C_N$ (Cu, Cr, Au), interface states $C_I$, and
extended states $C_E$ which have a probability amplitude of approximately the
same size in all regions. 
To emphasize the contribution of the different classes of
eigenstates the total conductivity (\Eq{eq_spincond}) was projected to the
layers $\nu$ in the supercell according to \Mcite{zahn02} 
\begin{eqnarray}
\sigma^\sigma(\nu)=    \frac{e^2}{V}
\sum\limits_k \delta
(E_k^{\sigma}-E_F){\bf v}_{k,x}^{\sigma} {\bf v}_{k,x}^{\sigma}
\left|c_k^\sigma(\nu)\right|^2
\; ,
\label{eq_cond_split}
\end{eqnarray}
with $c_k^\sigma(\nu)$ the expansion coefficient of the Bloch eigenstate ($k$,
$\sigma$) at the site $\nu$ in the supercell. The normalization of the Bloch
state to the unit cell yields $\sum_\nu \left|c_k^\sigma(\nu)\right|^2 = 1$.
In addition, the site dependent conductivity $\sigma^\sigma(\nu)$ was split
into the contributions of the 4 typical classes of eigentstates which are shown
in \Fig{fig_sig_local}. 
%
%
It is evident, that a large contribution of the CIP conductivity is carried by
the QWS in the magnetic and nonmagnetic layer. Especially, in the minority
channel of the Co/Cu system and the majority channel of the Fe/Cr system the
conductivity is dominated by contributions of the magnetic quantum well
states. By considering only a few types of defects in the sample the
conductivity is dominated by one spin direction in most cases. Assuming more
types of defects and taking into account the effect of self averaging this
tendency is reduced. 

%
%
\begin{figure}
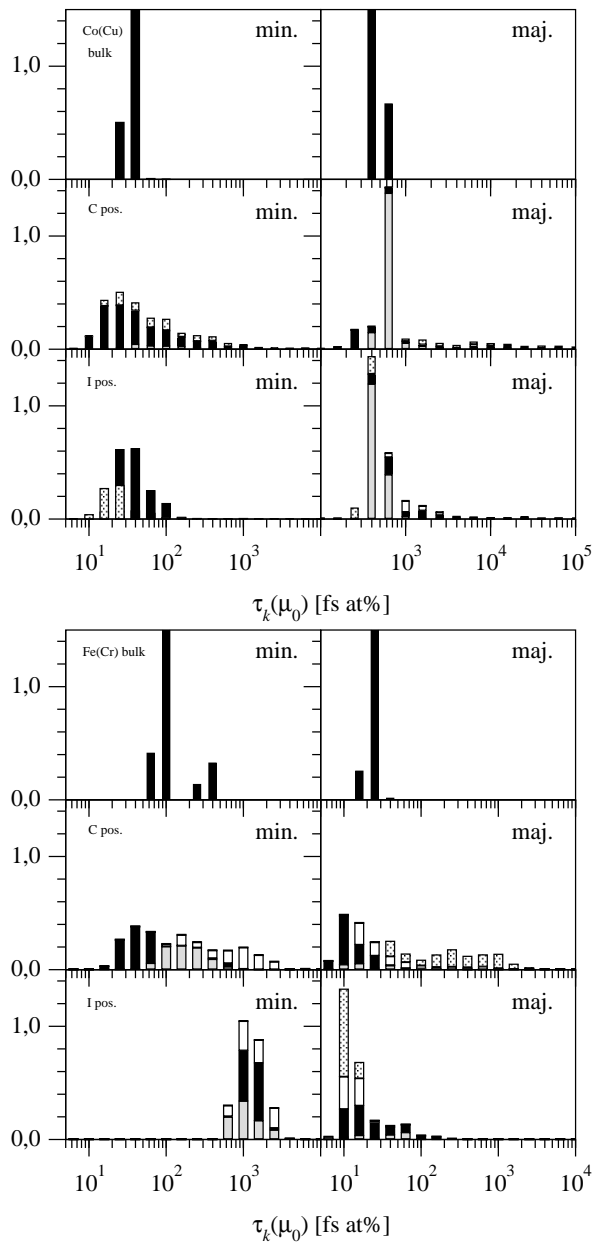

\Mi=.9\linewidth
\includegraphics[width=\Mi,clip=true]{fig7a.eps}\\
\includegraphics[width=\Mi,clip=true]{fig7b.eps}
\caption{
histogram of spin-dependent, anisotropic relaxation times of
$Co_9Cu_7$ (top panel), and
$Fe_9Cr_7$ (bottom panel) for P alignement, defects in the magnetic layer
at interface position (bottom), center position (middle), 
and defect in bulk (top),
the contribution of typical classes are marked by 
grey - extended, black - Co/Fe, white - Cu/Cr/Au, and 
light grey - interface states
}
\label{f:tau-hist}
\end{figure}
%
Due to the quantum size effects the relaxation times show a strong
variation for the states at the Fermi level which determine the
conductivity. 
All Bloch states with a nearly zero probability amplitude
at the impurity site undergo a weak scattering and cause extremely large
relaxation times. 
The state
dependent relaxation times are distributed over several orders of
magnitude, especially for defects inside the
metallic layers (see \Fig{f:tau-hist}).
This is a new effect that does not occur in bulk
systems.
The panels show the relative amount of relaxation times $\tau_k$ for the
states at the Fermi level for the Co/Cu system (top viewgraph) and the Fe/Cr
system (bottom panel). 
The spin resolved histograms for Cu defects in bulk Co are given for
comparison (topmost subpanel). 
The remaining subpanels represent relaxation times for Cu defects in the center of the
Co layer and for Cu defects at the Co/Cu interface.
The color of the bars labels the character of the eigenstates: 
extended multilayer states are shown in dark grey,
QWS confined to the magnetic layer are shown in black. 
QWS confined to the nonmagnetic layer are given in white, and
interface states are given by light grey bars.
For defects inside the magnetic layer the maximum of the distribution coincides
with that in the bulk material. In addition, a long tail for high values occurs
caused by states which have a small probability amplitude at the defect
position, e.g. quantum well states in the nonmagnetic layer or interface
states. This is best seen in the middle subpanel for Cr defects in the center
of the Fe layer for the Fe/Cr multilayer. Quantum well states confined to the
Fe layer have a large probability amplitude at the defect position and as a
result smaller relaxation times than in the bulk system. Interface states with
a tail penetrating the Fe layer are scattered on an intermediate level and Cr
quantum well states with the lowest probability amplitude at the defect
position are scattered weakly.

The states with large relaxation times although not numerous
are highly conducting and nearly provoke a short circuit. 
This is the case for Co impurities in the Cu layer for the Co/Cu multilayer,
compare \Fig{f:tau-hist}, top panel, and for Fe defects in the Au layer for
the Fe/Au system. 
This effect is mainly obtained
for impurities in the center of the layers and is related to the fact that
in-plane transport is mostly driven by quantum well states \cite{zahn98}.
%
This peculiar behavior of conductivity is in
agreement with the results of Blaas et al.\ \cite{blaas99a}
who found higher resistivities for Co/Cu multilayers with
interdiffusion restricted to the interface layers than
for alloying with Cu atoms in the Co layers.

%
%
\begin{figure*}
  \Mi=.3\linewidth
  \setlength{\unitlength}{\Mi}%
  \psset{unit=\Mi}%
  \begin{picture}(1,1.52)(0,0)%
  \rput[bl](0,0){\includegraphics[width=\Mi,clip=true]{fig2a.eps}}
  \rput(.495,.440){%
\includegraphics[width=.50\Mi]{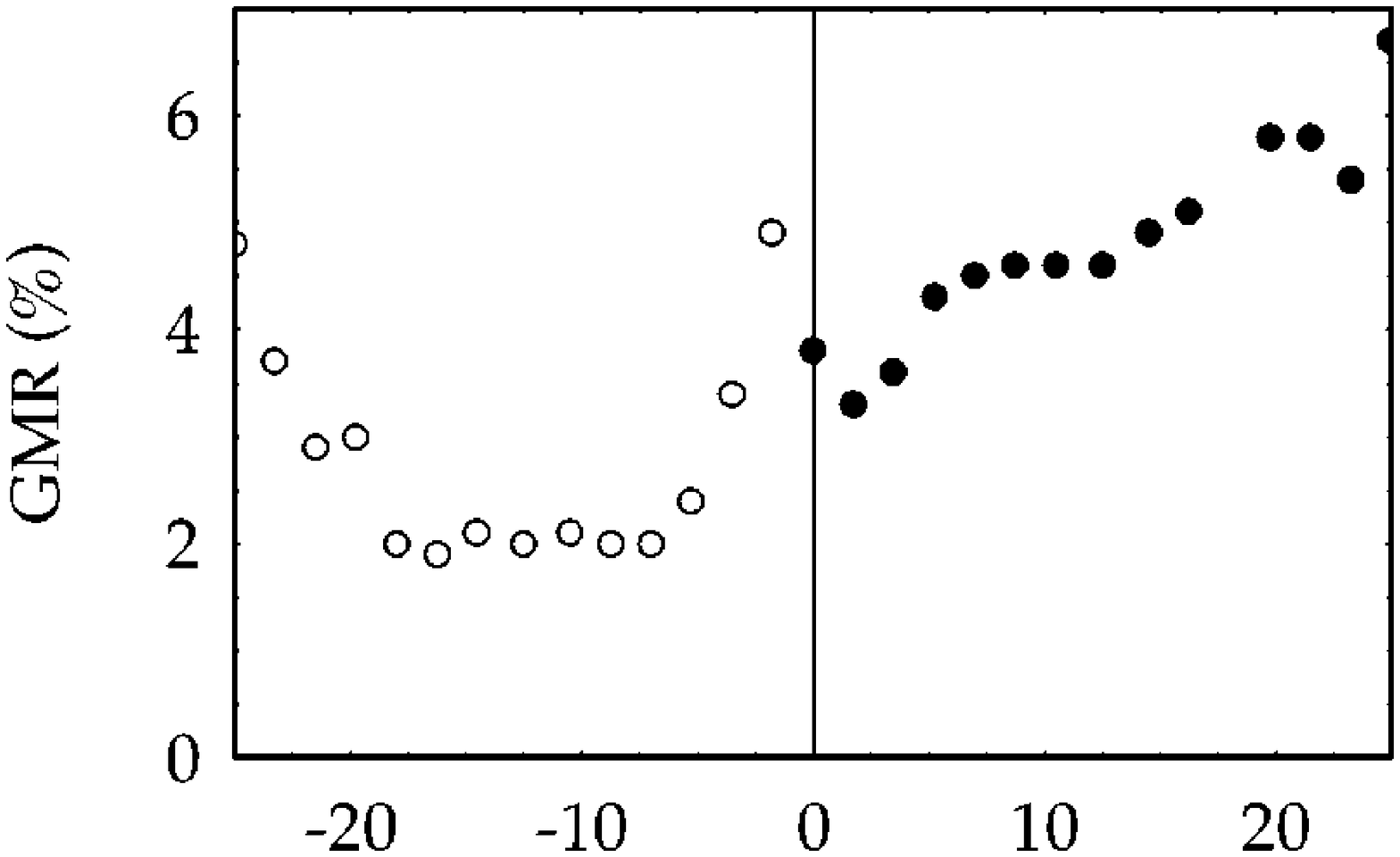}}%
  \end{picture}
\includegraphics[width=\Mi,clip=true]{fig2b.eps}
\includegraphics[width=\Mi,clip=true]{fig2c.eps}
  \caption{Dependence of GMR in \CoCu, \FeCr, and \FeAu on the position of   'self' impurities.
  a-c) Assuming scattering at the inserted $\delta$-layer only,
  d-f) assuming $\delta$-layer scattering ($50\%$) and interface
  scattering ($50\%$). The thin dashed lines marks the GMR ratio
  assuming interface impurities only, and the thick dashed line gives
  the GMR assuming isotropic relaxation times.
  the inset in panel (d) shows the experimental result from
  Ref.~\cite{marrows01} 
   }
  \label{fig_gmr_all}
\end{figure*}
For comparison with experiments we have to mention that the large
absolute values would hardly be obtained experimentally
since they correspond to idealized samples with perfectly flat interfaces
and defects at well defined positions in the superlattice. 
As soon as we consider
an overall distribution of defects in the multilayer the highly conducting 
channels are suppressed. 
The general trend, however, survives.
This phenomenon of highly conducting electrons confined to one
layer of a multilayer structure is called electron wave guide or
channeling effect \cite{butler96,stiles96} and
was also experimentally verified \cite{dekadjevi01}. 
\par
Structural investigations of Co/Cu multilayers on an
atomic scale \cite{larson98,schleiwies01}
gave evidence that most of the structural imperfections appear next to
the interfaces. To investigate the influence of
more than one type of scattering centers in one sample a simplified
defect distribution was assumed. 
In addition to
the specific $\delta$-layer defects of the magnetic layer material in the
nonmagnetic interface atomic
layer and defects from the nonmagnetic material in the magnetic interface
layer are 
considered to simulate an intermixed region at the interface. For
the concentration weights $x(\mu)$ entering \Eq{eq_tau_mix} we
choose $25\%$ for defects in both of the interface layer and $50\%$ for
defects in the 
$\delta$-layer and the resulting conductivities are shown in
\Fig{fig_sig_all}(b),(d),(f), respectively. 
First, an overall reduction of the resistivities is obtained, caused by the
distribution of defects at different positions and the resulting higher
probability that also quantum well states are scattered at one or the other
type of defect. According to \Eq{eq:tau-mix} the defect with the highest
scattering rate $1/\tau_k$ dominates the total relaxation time
$\tau_k$. 
\section{Giant magnetoresistance}
The GMR ratios derived from the conductivities in
\Fig{fig_sig_all} are shown in \Fig{fig_gmr_all}.
Assuming scattering centers in the {\DL} only (\Fig{fig_gmr_all} (a),(c),(e))
huge GMR ratios are obtained  especially for Cu defects in the Co
layer. Introducing additional interface scattering with a weight of 50\% causes strongly
reduced values (\Fig{fig_gmr_all}(b),(d),(f).
The thin dashed line in the lower panels of \Fig{fig_gmr_all} is the GMR
ratio caused by interface scattering.
This value would correspond to the reference value in the experiments of Marrows
and Hickey without
{\DL} \cite{marrows01}. 
The thick dashed line in \Fig{fig_gmr_all} gives the GMR value
obtained with the assumption of a constant relaxation time without any
spin or state dependence (intrinsic GMR). In comparison to this case of isotropic
scattering the insertion of an additional {\DL} increases GMR, mostly at
the interfaces.

%
Comparing the trend of GMR an excellent agreement with the experiment
(inset in \Fig{fig_gmr_all}) is obtained for the Co/Cu system. 
To our knowledge, up to now similar experimental investigations of the GMR
dependence on the defects position are not available for Fe/Cr
multilayers.
The importance of the interface scattering to obtain a large GMR
effect in Fe/Cr was pointed out by several authors
\cite{davies96,schad98,cyrille00}. 
The weighting factors for the different scattering mechanisms are
derived from the structural investigations by Davies et
al. \cite{davies96}. We choose $0.45$ for Cr defects in the Fe
interface atomic layers and $0.05$ for Fe defects in the Cr interface
layer. The contribution from the {\DL} is fixed to $0.5$ as in the case
of the Co/Cu multilayer. The most striking feature in comparison to
Co/Cu is the reduction of the GMR ratio by introducing Cr defects in
the Fe layers
as shown in
\Fig{fig_gmr_all}b) and e). 

We still
have to mention, that the calculated values are two orders of magnitude
larger than the experimental ones. The reason is the restriction to
substitutional point defects. In addition to these much more scattering
mechanisms are active in real samples. 
Assuming self-averaging the results could be corrected towards the experimental
ones by an additional spin- and state-independent relaxation time
$\tau$ (thick dashed line in fig. 2d-f) \cite{zahn98}. 

In contrast to
\Mcite{zahn98s} the present results were obtained assuming the above
described impurity distribution only and are focussed on the impurity
scattering rates only. 
Another difference to the experimental setup in \Mcite{marrows01s}
is the considered geometry.
The experimentally investigated samples have been Co/Cu/Co spin valves
grown on a buffer layer and protected by a cap layer. As a
consequence the GMR ratios are nearly symmetric as a function of the
impurity position in the Cu layer but asymmetric for defects in the Co
layer. The calculations are performed in supercell geometry which is
reflected in the symmetry of the results with respect to the defect
position in both layers, Cu and Co. A possible influence of superlattice
effects in metallic multilayers was shown to be negligible \cite{erler01}. 
\par
\begin{figure}
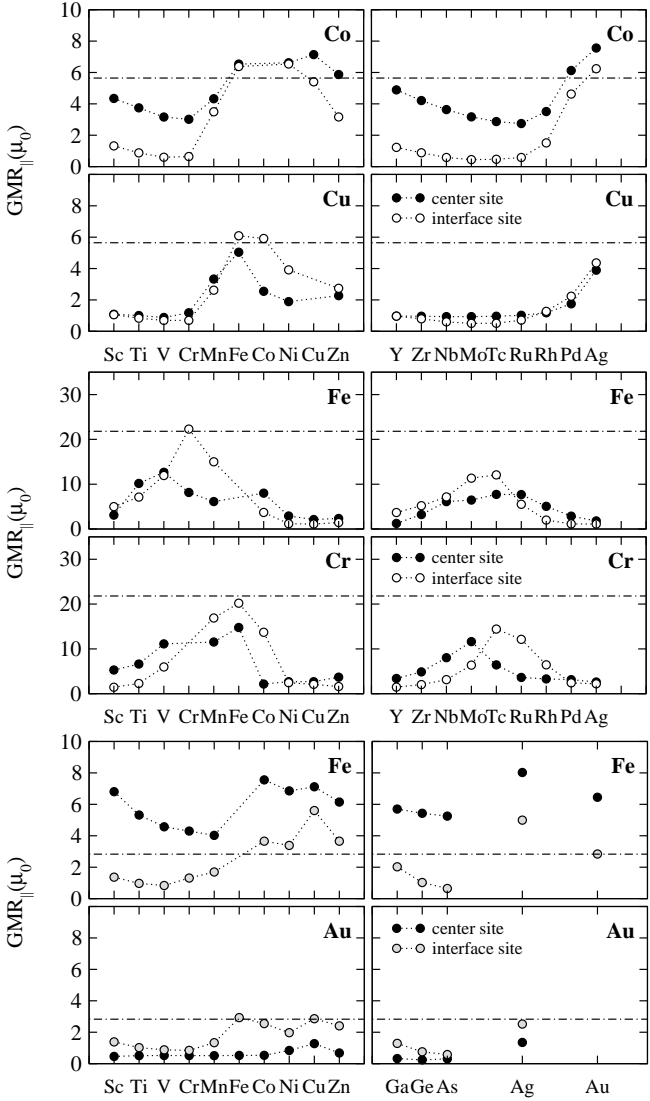

  \Mi=.99\linewidth
\includegraphics[width=\Mi,clip=true]{fig3a.eps}\\
\includegraphics[width=\Mi,clip=true]{fig3b.eps}\\
\includegraphics[width=\Mi,clip=true]{fig3c.eps}
  \caption{Dependence of the GMR on the impurity {\DL} 
   doped with different transition and sp-metals and the position in the
   in \CoCu (top panel), \FeCr(middle panel), and \FeAu(lower panel);
   the label of each subpanel denotes the layer where the {\DL} is inserted;
   closed symbols: positions in the center, and open symbols mark the
   interface positions
   }
  \label{fig_gmr_def}
\end{figure}
\Fig{fig_gmr_def} compiles the trend of GMR caused by 3d-,4s- (left column) and
4d-,5s-transition metal impurities (right column) as a function of position
in the magnetic layer. 
The CIP-GMR ratios are given for the corresponding defects in the middle of
the magnetic and nonmagnetic layer, respectively, marked by the closed
symbols. The open symbols correspond to a position of the {\DL} at the
interface. The horizontal dashed lines indicate the case of interface
scattering only. This means we consider a {\DL} of the magnetic material in the
nonmagnetic interface atomic layer and vice versa. The thick dashed lines give
the results assuming a spin and state independent relaxation time. 
\par
\par
%
%
In a previous work \cite{zahn98} we used a simplified model to
describe the scattering. We assumed $\delta$-peak like
potential perturbations characterized by a spin-dependent scattering strength
$t^\sigma$. Comparing with these results one can state the following. 
The reader should note that the order of Co and Cu in \FIG{2} in
ref.~\cite{zahn98} is reversed in comparison to
\Fig{fig_gmr_all}a) and d) of this work.
The insertion of a {\DL} in the Co layer enhances the GMR in comparison
to the undoped case in both models.
The differences of the results are caused by the
approximation used for the scattering strength. In this work the
scattering potentials were determined self consistently also for a
region around the defects.
Evaluating \Eq{eq_tkk} the Born series expansion of the
scattering operator $T$ in terms of the single site t-matrix and all
multiple scattering contributions are included completely.

Comparing the influence of 3d defects at the interface with that
caused by an ordered interface alloy \cite{zahn01} a similar trend for
the GMR is obtained. One should compare \FIG{5} (left column) in
ref.~\cite{zahn01} and the GMR values for the position $x=0$ of
the {\DL} in \Fig{fig_gmr_def}. For the lighter 3d elements up to Mn
the GMR is lowered with the interface alloy, whereas for heavier
elements the GMR is maintained or even increased.
\par
All the calculations are carried out for the limit of low defect
concentration. 
That means the scattering at different defects is
treated independently. 
For a typical concentration of 1\% we obtain
an imaginary part of the self energy of the order of
$10^{-3}eV$. This is small in comparison with the typical band width
in the transition metals. 
In contrast to the work of Tsymbal et al. \cite{tsymbal96} the
contributions to the conductivity arising from interband transitions
are expected to be small.
\par
\section{Summary}
%
In conclusion, the self-consistent calculation of the scattering
properties and the improved treatment of the Boltzmann transport
equation including vertex corrections provide a powerful tool for
a comprehensive theoretical description and a helpful insight into
the microscopic processes of CIP-GMR. The experimentally found
trends concerning the doping with various materials at different
positions in the magnetic multilayer could be well reproduced
which means that spin-dependent impurity scattering is the most
important source of GMR. The theoretical results show furthermore
that interface scattering caused by intermixing plays a crucial
role and has to be taken into account in any system under
consideration. Selective doping of the multilayer with impurities
in specific positions causes variations of GMR which could be well
understood by the modulation of spin-dependent scattering due to
quantum confinement in the layered system and by the spin
anisotropy $\alpha$.
\begin{acknowledgments}
Financial support by the DFG (FG 404) and BMBF contract 13N7379 is
kindly acknowledged.
\end{acknowledgments}
%
%
%
%

%
\end{document}